\documentclass[sigconf]{acmart}
\usepackage{graphicx}
\usepackage{subcaption}
\usepackage{tabularx}
\usepackage[many]{tcolorbox}   % Required for all box styles (also loads xcolor and tikz)
\usepackage{xcolor}            % For color definitions
\AtBeginDocument{%
  }

\definecolor{main}{HTML}{5989cf}
\definecolor{light}{HTML}{e5f0fd}
\definecolor{sub}{HTML}{cde4ff}
\definecolor{frame}{HTML}{5b7ae0}
\newtcolorbox{boxF}[1]{
    fonttitle=\bfseries,
    colback = sub,
    enhanced,
    boxrule = 1.5pt, 
    colframe = frame, % base frame color
    title=#1
}

\newtcolorbox{boxC}{
    colback = light, % background color
    boxrule = 0pt  % no borders
}

\acmConference[PASC 2026]{The Platform for Advanced Scientific Computing}{June 29--July 1, 2026}{Bern, Switzerland}
\begin{document}

\title{Exploring Sustainability in Scientific Software through Code Quality \& Test Coverage Metrics}

\author{Sheikh Md. Mushfiqur Rahman}
    \affiliation{%
        \institution{University of Tennessee}
        \city{Knoxville}
        \state{TN}
        \country{USA}
}
\email{srahma14@vols.utk.edu}

\author{Gregory R. Watson}
    \affiliation{%
        \institution{Oak Ridge National Laboratory}
        \city{Oak Ridge}
        \state{TN}
        \country{USA}
}
\email{watsongr@ornl.gov}

\author{Nasir U. Eisty}
\affiliation{%
	   \institution{University of Tennessee}
	   \city{Knoxville}
	   \state{TN}
	   \country{USA}
}
\email{neisty@utk.edu}

\renewcommand{\shortauthors}{Rahman et al.}

\begin{abstract}
\textit{\textbf{Context:}} Scientific open-source software (SciOSS) plays a foundational role in research and engineering, yet its long-term sustainability has often been overlooked and remains a significant concern. 
\textit{\textbf{Objective:}} This study investigates the long-term sustainability of SciOSS through code and test quality metrics.
\textbf{\textit{Method:}} We analyze CASS Software Portfolio projects, classifying them by sustainability and comparing their code structure, test coverage, and links between code quality and testing across the dataset. 
\textbf{\textit{Results:}} Sustainable projects show higher, more consistent test coverage and clearer code–test correlations, while unsustainable ones show weaker patterns. Overall, test coverage is low in scientific software, and high complexity and coupling reduce testability. 
\textbf{\textit{Conclusion:}} In this study, we present a practical, data-driven approach for assessing sustainability in scientific software, offering a foundation for evaluating long-term software health and supporting future efforts in quality assurance and sustainability monitoring.

\end{abstract}





\maketitle
% \copyrightnotice

\section{Introduction}
%Open-source software (OSS) has significantly democratized software development, making it accessible to a broader audience. The sustainability of OSS projects is critical to the infrastructure of our digital society. Successful and sustainable OSS projects dominate the Internet and are reportedly used in 98\% of enterprises~\cite{pcmagSurveyPercent}. Many developers contribute to OSS to implement specific features, support meaningful causes, or enhance their skills.

%However, aside from a few blockbuster projects, many OSS initiatives, especially smaller or nascent ones, struggle to achieve and maintain long-term sustainability. In fact, over 80\% of OSS projects eventually become inactive~\cite{schweik2012internet}. The consequences of a widely-used OSS project falling off a sustainable trajectory can be severe, as demonstrated by recent global cybersecurity incidents like the log4j vulnerability~\cite{apacheApacheLog4j}.

Scientific software, broadly defined, refers to software developed for scientific purposes, primarily to improve our understanding of real-world phenomena or to enable predictive modeling of such processes~\cite{kanewala2014testing,mitreCVE202144228}.
In a research context, such software may be developed with limited knowledge of software requirements (specifications and features of software), industry-standard software design patterns~\cite{gamma1994design}, good coding practices (e.g., using descriptive variable names), version control, software documentation, automated testing, and project management practices (e.g., Agile)~\cite{carver2022survey,reinecke2022critical}. This leads to the creation of source code that is not well-structured, is not easily (re)usable, is difficult to modify and maintain, has scarce internal documentation (code comments) and external documentation (e.g., manuals, guides, and tutorials), and has poorly documented workflows~\cite{nyenah2024software}.

Scientific software that suffers from these shortcomings is likely to be challenging to sustain and have severe drawbacks for scientific research. For example, it can impede research progress, decrease research efficiency, and hinder scientific progress, as implementing new ideas or correcting mistakes in code that is not well-structured is more difficult and time-consuming~\cite{nyenah2024software}. In addition, it increases the likelihood of erroneous results, thereby reducing reliability and hindering reproducibility~\cite{Reinecke2022-vx}. These harmful properties can be averted, to some extent, with sustainable scientific software.

There are various interpretations of the meaning of sustainable scientific software. Anzt et al.~\cite{anzt2021environment} define sustainable scientific
software as software that is maintainable, extensible, and
flexible (adapts to user requirements); has a defined software
architecture; is testable; has comprehensive in-code
and external documentation, and is accessible (the software
is licensed as open source with a digital object identifier
(DOI)). Katz~\cite{katz2022research} views scientific software sustainability as the process of developing and maintaining scientific software that continues to meet its purpose over time.

Existing research has explored the sustainability of SciOSS by examining factors such as early development behavior and project activity~\cite{xiao2023early,han2024sustainability}, authorship and contributor participation~\cite{avelino2019abandonment}, social and technical networks~\cite{yin2021sustainability}, and project-level metrics~\cite{stuanciulescu2022code,nyenah2024software}. Building on this foundation, our study adopts a data-driven approach to investigate sustainability in SciOSS through the lens of structural code metrics and test suite coverage. We curated and analyzed a set of C/C++ SciOSS projects stewarded by the Consortium for the Advancement of Scientific Software (CASS)~\cite{cass_community_software_2025}, which is an initiative that emphasizes community engagement, reproducibility, and engineering rigor. We labeled each project as either sustainable or unsustainable using two complementary and well-established classification methods: one based on long-term commit activity, and the other based on the presence and continuity of core contributors, known as Truck Factor (TF) developers.

By comparing these two groups, we examine how structural complexity, modularity, cohesion, coupling, and test coverage vary between sustainable and unsustainable scientific software. We also explore whether these metrics exhibit meaningful correlations, particularly in sustainable projects, and assess whether stricter definitions of sustainability lead to more reliable distinctions in software quality. Our findings offer insights into the relationship between internal code quality and sustainability, helping to inform better practices for developing and maintaining scientific software. Our study is guided by the following research questions:
\begin{itemize}
    \item \textbf{RQ1: How do structural code metrics and test coverage differ between sustainable and unsustainable SciOSS?} Scientific software is often developed without rigorous engineering practices, leading to poor maintainability and increased risk of errors . While prior studies have primarily focused on social or activity-based indicators (e.g., contributors, commits), there is limited understanding of how internal code properties, such as complexity, modularity, and test coverage, relate to sustainability. This research question addresses that gap by investigating whether measurable structural and testing characteristics can distinguish sustainable projects from those at risk. Establishing such differences is important because it enables the identification of concrete, actionable indicators that developers and stakeholders can monitor to improve long-term software health.
    \item \textbf{RQ2: Does using a stricter threshold for sustained activity improve the reliability of sustainability classifications based on test quality metrics?} As shown in prior work and reinforced in this study, varying criteria (e.g., activity-based vs. contributor-based) can produce contradictory interpretations of quality indicators like test coverage. This research question is important because it examines whether refining the definition of sustainability by setting stricter activity thresholds reduces ambiguity and improves the reliability of empirical findings. Without such validation, conclusions drawn about sustainability may be misleading or dependent on arbitrary labeling choices. By evaluating the impact of stricter thresholds, this RQ contributes to methodological rigor and helps establish more reliable benchmarks for future sustainability research.
    \item \textbf{RQ3: Is there any strong relationship between code metrics and test coverage in scientific software overall?} It remains unclear whether intrinsic code properties, such as complexity, coupling, and cohesion, systematically influence testability. This research question is important because it seeks to uncover fundamental relationships between code structure and testing effectiveness across projects. Identifying such relationships can provide deeper insights into why certain software systems are harder to test and maintain. Moreover, understanding these links enables the use of structural metrics as early indicators of testing challenges, offering a scalable way to assess and improve software quality in large scientific codebases.
\end{itemize}

By answering these questions, we aim to shed light on the structural and testing-related attributes that distinguish sustainable scientific software from those at risk of becoming unsustainable. Our findings contribute to a better understanding of the engineering characteristics that underpin sustainable scientific software and offer actionable insights for developers, maintainers, and stakeholders seeking to improve the longevity and reliability of SciOSS.
\section{Related Work}
\citet{stuanciulescu2022code} investigated how code, process, and quality metrics correlate with the sustainability of Apache Software Foundation Incubator (ASFI) projects. By comparing over 200 graduated (sustainable) and retired (unsustainable) projects, the authors find that factors like contributor activity, commit patterns, and code complexity significantly influence project outcomes. 
% The study uses regression modeling to show that these metrics can help predict sustainability early in a project's lifecycle.

\citet{yin2021sustainability} presented a method to predict the sustainability of Apache Incubator projects by analyzing their monthly socio-technical metrics derived from commits and email interactions. Projects are labeled as sustainable if they graduate and unsustainable if they retire. A time-series LSTM model trained on these features achieves 93\% accuracy within 8 months of incubation. The authors also apply LIME to interpret the model and validate its predictions through real-world case studies. 

\citet{xiao2023early} proposed a machine learning model to predict the long-term sustainability of open-source GitHub projects early in their lifecycle, using behavioral signals such as founder history, development activity, community engagement, documentation, and popularity from the first few months. Projects are labeled as sustainable if they remain active for two years with a median monthly commit rate above a threshold. They found  that stricter thresholds improve the model's classification accuracy by reducing label noise.

\citet{han2024sustainability} proposed a forecasting approach that uses LSTM (Long Short-Term Memory) neural networks to predict whether deep learning packages on GitHub will remain active or become dormant. The model is trained on time-series data capturing monthly developer activity and collaboration patterns. In addition to making accurate predictions, they explored how activity trends can identify early signs of decline in otherwise sustainable-looking projects. 

\citet{valiev2018ecosystem} investigated what makes OSS projects in the Python Package Index (PyPI) ecosystem sustainable by combining large-scale quantitative analysis with developer interviews. The authors define project sustainability as continued maintenance and label projects as dormant if they average less than one commit per month in the year before their last commit. Using logistic and linear regression models, they evaluate how ecosystem-level factors, such as dependency centrality (PageRank), backporting, and organizational support, affect project survival, revealing that ecosystem position plays a significant role in sustainability. 

\citet{avelino2019abandonment} studied how OSS projects survive or become abandoned after losing their core developers (Truck Factor developers). It introduces the TF Developer Detachment (TFDD) concept and compares surviving vs. non-surviving projects using metrics like number of developers, commits, files, and project age. Statistical tests (Mann–Whitney U, Cliff’s Delta, Benjamini–Hochberg) reveal that surviving projects tend to be more active, older, and attract new key contributors.

Complementary to the above prediction-oriented studies, prior work on sustainability has also emphasized multi-signal repository stability models (often referred to as CSI frameworks), which integrate commit activity, contributor retention, responsiveness, dependency health, and governance signals into composite sustainability indicators~\cite{coelho2017modern,goggins2021open,oriol2023comprehensive,dias2021makes}.

Unlike prior studies that predict sustainability in general-purpose OSS or specific ecosystems (e.g., ASF Incubator, PyPI, GitHub) using process, social, and activity metrics, our study emphasizes the relationship between code-level structural \& test metrics with sustainability in SOSS. Using projects from the CORSA Catalog, we combine detailed static code metrics (size, complexity, cohesion, coupling, documentation, function calls) with dynamic test coverage measures (line, branch, function) to investigate how internal code quality relates to long-term sustainability.
\section{Methodology}
\subsection{Project Selection Criteria}
% Unlike prior work that typically relies on large-scale automated mining of general-purpose open-source projects from GitHub, our study focuses specifically on scientific open-source software with verified licenses, active community engagement, and substantial engineering complexity. To this end, 
We curated a set of open-source scientific software projects written in \textbf{C and C++}, drawn from the CASS Software Portfolio~\cite{casscatalog}. 
All selected projects are publicly hosted (e.g., on GitHub or institutional repositories), available for cloning and local builds. The CASS Portfolio includes tools widely used across domains such as artificial intelligence, mathematics, physics, performance and correctness analysis, programming systems, and scientific data management and visualization, with a strong emphasis on community relevance, reproducibility, and long-term sustainability.

%Despite curating candidates from the CORSA Catalog, 
Several projects in the CASS Portfolio proved infeasible to build in a constrained research environment. In practice, we encountered (i) high memory and link-time requirements (often >16–32 GB RAM), (ii) platform-specific or privileged dependencies (e.g., CUDA/ROCm drivers, MPI stacks, legacy toolchains), and (iii) brittle or undocumented build scripts that failed under coverage instrumentation. We standardized builds in Docker with GCC/Clang and gcovr, but excluded repositories that could not configure or compile with \texttt{--coverage -O0 -g}, required specialized hardware, or lacked runnable tests. This feasibility filter reduces the sample size but improves the reproducibility and internal validity of our measurements.
To ensure compatibility with our instrumentation and analysis workflow, we restricted our selection to projects that satisfied the following criteria, enabling both static analysis (e.g., AST parsing) and dynamic analysis (e.g., test coverage):

\begin{figure}[h]
  \centering
  \includegraphics[width=\linewidth]{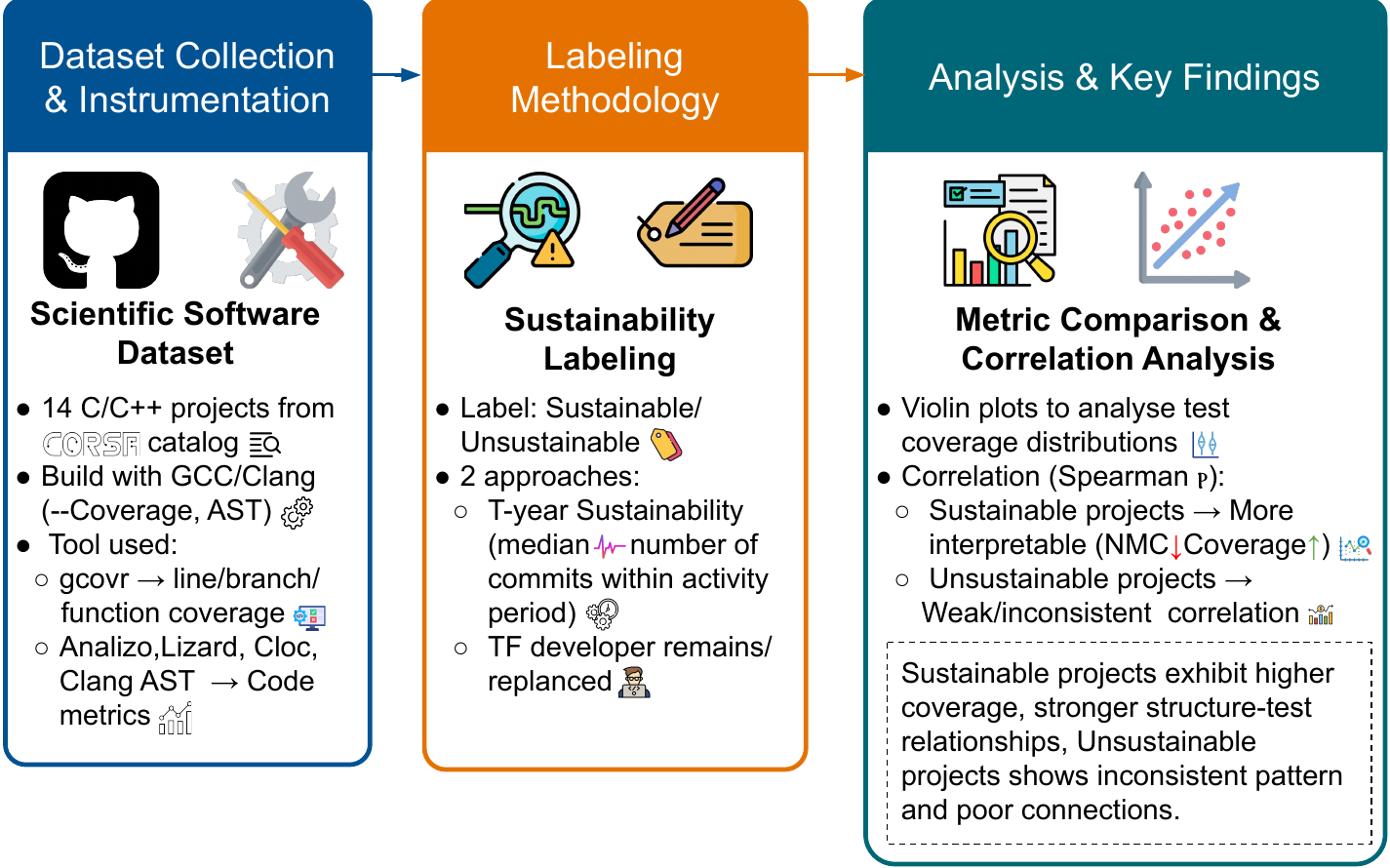}
  \caption{Overview of The Approach \& Key Findings.}
\end{figure}

\begin{itemize}
\item The project must be written in \textbf{C and/or C++}, and hosted in a publicly accessible repository to allow cloning, local builds, and transparent reproducibility.

\item It must use a supported build system such as \textbf{GNU Make}, \textbf{CMake}, or \textbf{GNU Autotools} (e.g., \texttt{configure}, \texttt{autogen.sh}) to ensure reproducible builds, correct dependency resolution, and seamless integration of coverage instrumentation flags across multiple compilation units.

\item It must \textbf{build successfully} using \texttt{gcc} or \texttt{clang} with coverage instrumentation flags (\texttt{--coverage -O0 -g}). These flags enable line- and branch-level tracking by injecting profiling logic into the compiled object files. Each compilation unit must be instrumented individually to produce \texttt{.gcno} and \texttt{.gcda} files to be processed into coverage summaries.

\item %In addition to \texttt{gcc}-based coverage, 
It must be \texttt{clang} compatible to extract \textbf{abstract syntax trees (ASTs)}.
% , which we used for computing fine-grained static metrics such as total number of function calls (NMC), number of internal method calls (NMCI), and number of external method calls (NMCE).

\item It must contain at least one \textbf{test executable} or driver that runs successfully (i.e., produces “green” test results), which is essential to generate runtime coverage data. Tests must exercise enough code paths to produce meaningful and analyzable \texttt{.gcda} files.

\item It must be compatible with \textbf{\texttt{gcovr}}~\cite{gcovr}, the coverage analysis tool we used to generate standardized line, branch, and function coverage reports. \textbf{\texttt{gcovr}} is a  lightweight command-line tool that wraps the GNU \texttt{gcov} utility. 
% It analyzes \texttt{.gcda} and \texttt{.gcno} files produced during test execution and generates both machine-readable reports (CSV, XML) and human-readable reports (HTML) suitable for inspection and post-processing.
Our pipeline ran \texttt{gcovr} from the build directory, using project-root anchoring (\texttt{-r}) and supporting both tabular and HTML outputs.
\end{itemize}

\subsection{Metrics}
To characterize the structural complexity, modularity, and documentation of each source file,  we collected a diverse set of static and dynamic code metrics. These metrics span multiple dimensions, including size, comments, function calls, complexity, cohesion, coupling, and test coverage, enabling a detailed structural analysis of both C and C++ codebases. These metrics reflect not only code volume and granularity, but also the extent of in-line documentation and function interface design. The full set of collected metrics is summarized in Table~\ref{tab:metrics}.

\subsubsection{\textbf{Size and Comment Metrics}}
We collected a set of size and comment-related metrics, applicable to both C and C++ projects. These metrics provide foundational insights into the structural volume and documentation practices of each source file. Metrics such as Lines of Code (LOC), Number of Methods (NOM), and Method Size (MMLOC, AMLOC) reflect the complexity and granularity of code at the function level. The Number of Attributes (NOA),  captures the density of field declarations and global variables, adding another layer to structural analysis. Complementing these are the documentation-related metrics, such as Lines of Comments (LOCCOM) and Comment Density (CD), which quantify the extent of in-line developer commentary. Together, these metrics characterize both the scale of implementation and the degree of self-documentation, which are important factors influencing maintainability and understandability across both C and C++ codebases.

Metrics related to static fields or bytecode instructions (e.g., NBI, NSTAF) were excluded as they are not directly applicable or consistently defined across C and C++.

\subsubsection{\textbf{Function Call Metrics Extraction}}

To analyze procedural \texttt{.c} files and extract function call-related metrics, we developed a static analysis script built around Clang's \texttt{-ast-dump} functionality. The script parses both the Clang-generated AST and the source code itself to identify and classify function definitions and invocations within each C source file. Specifically, we extracted Number of Method Calls (NMC), Number of Method Calls External (NMCE), and Number of Method Calls Internal (NMCI).

% \begin{itemize}
%   \item \textbf{Total function calls}: the number of \texttt{CallExpr} nodes found in the Clang AST, corresponding to all function invocations.
%   \item \textbf{Internal function calls}: function calls where the target function is defined within the same file.
%   \item \textbf{External function calls (NMCE)}: function calls to targets that are \emph{not} defined in the current source file.
% \end{itemize}

We identified function definitions through source-level pattern matching and detected function calls by analyzing the AST output. We extracted Callee names by scanning lines surrounding each \texttt{CallExpr} node. We considered a function call \emph{external} and thus counted it towards the NMCE metric if its target function does not match any function defined within the file under analysis. This strategy provides a lightweight approximation of inter-module coupling and behavioral complexity by treating each \texttt{.c} file as a logical unit similar to a class or module.

\subsubsection{\textbf{Complexity}}
Complexity metrics such as Weighted Methods per Class (WMC) and Average Method Complexity (AMC) are based on the cyclomatic complexity measure, which quantifies the number of linearly independent paths through a method’s control flow~\cite{mccabe1976complexity}. We also collected Response for Class (RFC)~\cite{chidamber1994metrics} which calculates the sum between the number of methods in the module and the number of functions called by each module function, and Structural Complexity (SC),  which quantifies how difficult a software module may be to maintain by combining two key design aspects: coupling and cohesion. Specifically, it is calculated as the product of CBO (Coupling Between Objects) and LCOM4 (Lack of Cohesion of Methods)~\cite{darcy2005structural}. A higher SC value indicates that a module is both tightly connected to other modules and internally less cohesive, which together suggest increased structural complexity and greater maintenance effort. %This approach follows the methodology proposed by Meirelles, based on Darcy et al.'s research~\cite{darcy2005structural}.

\subsubsection{\textbf{Cohesion}}
Cohesion refers to the degree to which the elements within a software module, such as statements in a method or functions within a file, are functionally related. In object-oriented programming, cohesion typically describes how well the methods within a class work together to fulfill a single responsibility. High cohesion is desirable, as it improves understandability, maintainability, and reusability. Conversely, low cohesion indicates that a module performs a wide range of unrelated tasks, making it harder to comprehend and maintain.

To quantify cohesion, several metrics have been proposed. One widely used metric is LCOM4~\cite{hitz1995measuring} (Lack of Cohesion of Methods, version 4). It models a module (e.g., a class or a C file) as an undirected graph where nodes represent subroutines (methods or functions), and edges connect nodes that either share access to at least one variable or directly call each other. The LCOM4 value corresponds to the number of connected components in this graph; higher values indicate lower cohesion, indicating that the module contains independent clusters of logic.

%Analizo~\cite{terceiro2010analizo} generalizes this metric beyond object-oriented classes: it treats C files as cohesive units by modeling functions as if they were methods in a class-like container. This allows cohesion to be consistently assessed across both C and C++ codebases, providing a uniform view of structural modularity.

\subsubsection{\textbf{Coupling}}
Coupling metrics characterize the degree of interdependence between different modules. Coupling Between Objects (CBO) calculates the number of calls to other modules. ACC counts how many other modules reference, import, or call functions from the given module.

\subsubsection{\textbf{Test-Quality Metrics}}

To assess the thoroughness of each project's test suite, we computed \textit{line coverage}, \textit{branch coverage}, and \textit{function coverage} using the \texttt{gcovr} tool~\cite{gcovr}. These metrics quantify how much of the code is exercised by the project's test suite, serving as a proxy for test effectiveness. 

We compiled each project using \texttt{gcc} or \texttt{clang} with coverage instrumentation flags (\texttt{$--$coverage -O0 -g}), executed the test suite, and then ran \texttt{gcovr} to extract coverage reports, which include line (\nolinkurl{line_percent}), branch (\nolinkurl{branch_percent}), and function (\nolinkurl{function_percent}) coverage.  

% We used these normalized percentage values reported by \texttt{gcovr}, including \nolinkurl{line_percent}, \nolinkurl{branch_percent}, and \nolinkurl{function_percent}, aas the measures of line, branch, and function coverage, respectively. 
Higher coverage values generally indicate that a larger portion of the program has been exercised by tests, increasing the likelihood of revealing faults (provided appropriate assertions are present). While coverage alone cannot guarantee correctness, it remains a practical and scalable proxy for test adequacy in large-scale studies.

\begin{table*}[t]
\centering
\small
\caption{Summary of Source Code and Test Coverage Metrics}
\renewcommand{\arraystretch}{1.2}
\begin{tabularx}{\textwidth}{p{6.3cm} X c}
\toprule
\textbf{Metric} & \textbf{Description} & \textbf{Ref.} \\
\midrule

\textbf{Size \& Comment Metrics} & & \\
Lines of Code (LOC) & Non-empty, non-comment lines of code & - 
\\
Number of Methods (NOM) & Total number of function definitions per file & - \\
Number of Branches (NOB) & Total number of branches in a file & - \\
Max Method LOC (MMLOC) & Lines in the longest function & \cite{meirelles2013monitoramento} \\
Average Method LOC (AMLOC) & Mean size of all functions in lines & \cite{meirelles2013monitoramento} \\
Average Number of Parameters (ANPM) & Mean number of parameters per function & \cite{meirelles2013monitoramento} \\
Number of Attributes (NOA) & Count of global variables and struct/class fields & - \\
Lines of Comments (LOCCOM) & Total comment lines (computed using \texttt{cloc}) & - \\
Comment Density (CD) & Ratio of comment lines to total lines (code + comments) & \cite{arafat2009comment} \\
\addlinespace
\hline
\textbf{Function Call Metrics} & & \\
% Static Functions & Functions declared with \texttt{static} keyword & [1] \\F
Number of Method Calls (NMC) & All function invocations (from AST \texttt{CallExpr} nodes) & \cite{terragni2020measuring} \\
Number of Internal Function Calls (NMCE)& Calls to functions defined in the same file & \cite{terragni2020measuring} \\
External Function Calls (NMCE) & Calls to functions defined in other files & \cite{terragni2020measuring} \\

\addlinespace
\hline
\textbf{Complexity Metrics} & & \\
Weighted Methods per Class (WMC) & Sum of cyclomatic complexity of all functions & \cite{chidamber1994metrics} \\
Average Method Complexity (AMC) & Mean cyclomatic complexity of functions & \cite{tang1999empirical} \\
Response for Class (RFC) & Number of functions + number of distinct calls they make &  \cite{chidamber1994metrics} \\
Structural Complexity (SC) & Product of CBO and LCOM4: SC = CBO $\times$ LCOM4 & \cite{darcy2005structural} \\

\addlinespace
\hline
\textbf{Cohesion Metric} & & \\
 Lack of Cohesion of Methods (LCOM4)  & Number of disconnected components in a function-variable/call graph & \cite{terceiro2012understanding} \\

\addlinespace
\hline
\textbf{Coupling Metrics} & & \\
Coupling Between Objects (CBO) & Number of modules/functions referenced by the file &  \cite{chidamber1994metrics} \\
Afferent Coupling Count (ACC) & Number of modules that depend on this file & \cite{martin1994oo} \\

\addlinespace
\hline
\textbf{Test Coverage Metrics} & & \\
Line Coverage (LC) & Percentage of source lines executed during testing & - \\
Branch Coverage (BC) & Percentage of decision branches exercised by tests & - \\
Function Coverage (FC) & Percentage of defined functions called at least once & - \\

\bottomrule
\end{tabularx}
\label{tab:metrics}
\end{table*}

\subsection{Tools and Techniques for Data Collection}

To collect \textbf{code metrics}, we used a combination of open-source tools and custom analyzers:
\begin{itemize}
  \item To collect consistent source code metrics across both C and C++ projects, we used \textbf{Analizo}~\cite{terceiro2010analizo}. Originally designed for object-oriented metrics, Analizo generalizes its internal representation by treating each translation unit (e.g., a \texttt{.c} file) as a module, similar to a class in C++. This abstraction allows it to compute a common set of structural metrics, such as method counts, coupling, cohesion, and complexity (metrics such as \textit{CBO}, \textit{NOM}, \textit{RFC}, and \textit{LCOM4}) for both procedural and object-oriented code. Thus, it is well-suited for analyzing mixed-language scientific software projects.
  
  % \textbf{Analizo}, to compute classical object-oriented 
\item \textbf{Lizard}~\cite{martin2017c++}, a lightweight static analysis tool, was used to compute cyclomatic complexity and function-level metrics. It supports \texttt{C/C++} without requiring complete headers or preprocessing, making it practical for large scientific codebases. We used it to extract \textit{WMC} (total cyclomatic complexity per file) and \textit{AMC} (average complexity per function), based on its per-function analysis of non-comment lines, parameter count, and control flow complexity.

  \item \textbf{Cloc}~\cite{adanial_cloc}, to count the total number of \textit{comment lines} in each source file and Comment Density (CD). 
  
  \item \textbf{Custom Clang-based AST analyzers}, to extract additional metrics not directly supported by the above tools, such as internal versus external function calls.

 \item Finally, to compute \textbf{test-quality metrics}, we executed the test suite of each project and used \textbf{\texttt{gcovr}}~\cite{gcovr} to collect test coverage report.
\end{itemize}
 %After collection, we aggregated all metrics and stored them in structured CSV files for statistical analysis.

\section{Results}
\subsection{RQ1: How do structural code metrics and test coverage differ between sustainable and unsustainable SciOSS?}

Various indicators help evaluate an OSS project’s sustainability:  
1) \textit{Artifact-related indicators}, such as employing a modular and extensible architecture and maintaining high-quality documentation~\cite{stuanciulescu2022code};  
2) \textit{Economic indicators}, including sufficient funding, low total cost of ownership, and high added value~\cite{qiu2019going,valiev2018ecosystem,zhang2019companies};  
3) \textit{Supply chain indicators}, such as keeping dependencies up-to-date and having a large number of downstream dependents~\cite{valiev2018ecosystem,winters2020software};  
4) \textit{Development activity indicators}, including frequent commits, a well-staffed team, and regular contributions from newcomers~\cite{tan2020first,yin2021sustainability}.  

Among these indicators, sustained development activity stands out as a particularly reliable marker of sustainability, since software that lacks ongoing maintenance tends to degrade in usefulness over time—a notion reinforced by Lehman’s Law~\cite{lehman2005programs}. While prior work has used development activity as a proxy for both the success~\cite{midha2012factors} and sustainability~\cite{valiev2018ecosystem} of OSS projects, Chengalur-Smith et al.~\cite{chengalur2010sustainability} emphasize that true sustainability is characterized by consistent, long-term engagement, in contrast to success, which can be assessed at a single point in time. Thus, in this paper, we use sustained activity as the main proxy for studying OSS sustainability, following previous works~\cite{avelino2019abandonment,coelho2018identifying,ghapanchi2015predicting,valiev2018ecosystem}.

In this study, we adopted two complementary approaches to measure sustained activity ensure a more comprehensive labeling process, each of which has been previously used as a proxy for software sustainability~\cite{valiev2018ecosystem,xiao2023early}:

\begin{itemize}
    % \item A software is labeled as sustainable if it averaged more than 1 commit per month over the past 12 months prior to its most recent commit; otherwise, it is labeled as dormant!\cite{han2024sustainability}.
    \item A project is classified as exhibiting sustained activity over a duration of $t$ years if it satisfies two conditions: (1) it has recorded commits spanning more than $t$ years, and (2) its monthly median commit count meets or exceeds a threshold $k$. The parameter $k$ can be tuned, with larger values reflecting a more rigorous criterion for sustained project activity. Based on the optimal thresholds used in previous works~\cite{xiao2023early}, we set the the values $t$=2 and $k$=6.
    \item A project is labeled sustained if at least one Truck Factor (TF) (the minimal number of developers that the project depends on for its maintenance and evolution~\cite{williams2003pair}) developer remains active or new TF developers emerge after detachment~\cite{avelino2019abandonment}. 

\end{itemize}

After categorizing the projects using the two classification methods, we gathered code metrics from the source files of each project in both groups. We examined whether there were any differences in test coverage between the codebases of the two datasets.

\begin{figure}[h]
  \centering
  \includegraphics[width=\linewidth]{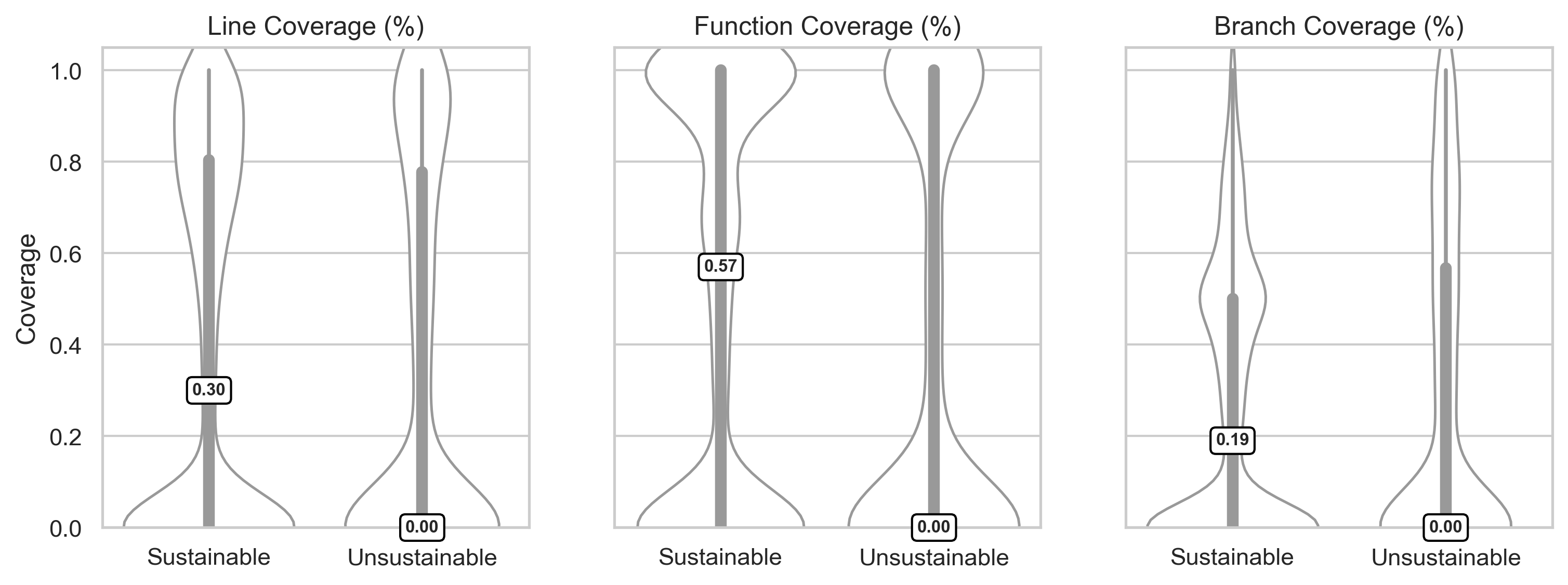}
  \caption{Coverage comparison based on $t$-year sustainability.}
\label{fig:coverage_tyear}
\end{figure}

First, we partitioned the dataset using the \textit{t-year sustainability criterion}, which labeled 8 projects as sustainable and 6 as unsustainable. In total, the sustainable projects contributed 5,579 files, while the unsustainable projects contributed 1,511 files. Figure~\ref{fig:coverage_tyear} summarizes the coverage distributions under this classification, where projects are labeled as sustainable if they maintain active development for a specified number of years. Under this classification, sustainable projects show clearly higher median values for line (0.30), function (0.57), and branch (0.19) coverage than unsustainable projects, which have a median of 0.00. This pattern suggests that projects with long-term activity tend to invest more consistently in testing infrastructure, possibly due to the demands of ongoing feature development, bug fixing, and maintenance. The stark contrast indicates that test coverage can serve as a distinguishing indicator of project longevity and sustained engagement.

\begin{figure}[h]
  \centering
  \includegraphics[width=\linewidth]{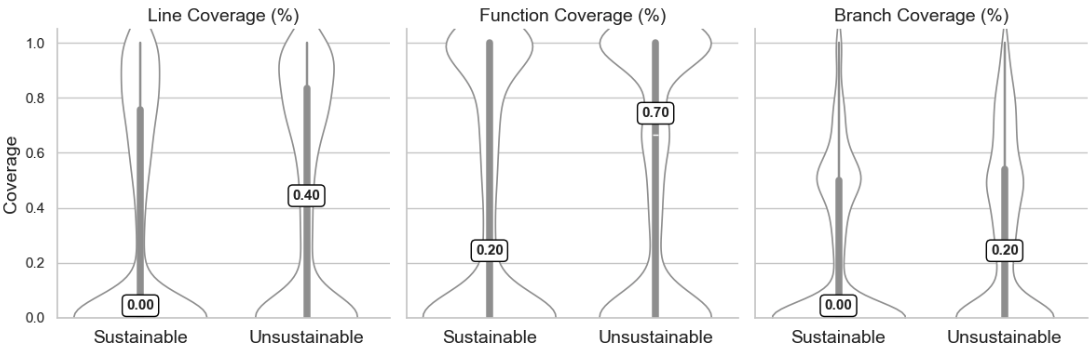}
  \caption{Coverage comparison based on TF classification.}
  \label{fig:coverage_tf}
\end{figure}

Then, we computed the Truck Factor (TF) following Avelino et al.~\cite{avelino2016novel}. Using this TF-based classification, 9 projects were labeled as sustainable and 5 as unsustainable.  In total, the sustainable projects contributed 3,305 source files, whereas the unsustainable projects contributed 3,785 files. Figure~\ref{fig:coverage_tf} summarizes the coverage distributions under this classification, where sustainability is inferred from the presence and continuity of multiple key contributors rather than prolonged activity alone. Interestingly, in this grouping, \textit{unsustainable} projects show higher median coverage values---0.40 for line, 0.70 for function, and 0.20 for branch---compared to \textit{sustainable} ones, which have noticeably lower values. This inversion may reflect a scenario where unsustainable projects were initially built and tested by a small, highly concentrated group of contributors but lacked broader community support for continued evolution. It underscores that while test coverage is important, it is not sufficient on its own to ensure sustainability if the developer base is fragile or narrow in scope.

\begin{boxC}
Sustainable projects show higher test coverage under the $t$-year criterion, but not under the TF-based classification, suggesting that coverage alone does not guarantee long-term sustainability.
\end{boxC}

Next, we computed the Spearman's correlation coefficient ($\rho$)~\cite{pearson2011comparison} for 45 ($15 \times 3$) pairwise combinations of structural and test-effort metrics. 
Spearman’s coefficient is a widely used non-parametric measure of correlation in related studies~\cite{bruntink2006empirical, toure2018predicting}. 
We did not use a parametric method such as Pearson correlation, as it assumes normally distributed data and linear relationships—assumptions that may not hold for our metric distributions. Instead, we opted for a non-parametric method that does not require such assumptions~\cite{hauke2011comparison}.

% follow a normal (Gaussian) distribution. Therefore, parametric correlation measures such as ``Pearson'' are inappropriate~\cite{hauke2011comparison}.
To verify non-normality, we applied the D’Agostino $K^2$ test~\cite{d1990suggestion}, which assesses both kurtosis and skewness to detect deviations from normality. 
% Alternative tests, such as the Shapiro–Wilk test, are not suitable due to the large number of observations (each distribution contains more than 5{,}000 data points)~\cite{razali2011power}. 
For all metric distributions, the $K^2$ test yielded $p$-values $\leq \alpha$ (with $\alpha = 0.05$), leading us to reject the null hypothesis of normality. Spearman’s coefficient ($\rho$) quantifies the degree to which two variables are associated with a monotonic function, i.e., an increasing or decreasing relationship~\cite{corder2009nonparametric}. 

The coefficient $\rho$ ranges from $-1$ to $+1$. A positive $\rho$ indicates that both variables increase together, while a negative $\rho$ suggests that one increases as the other decreases. A $\rho$ value close to zero implies little to no correlation between the two variables. 
% Figure~\ref{fig:heatmap} shows the heatmap of Spearman's coefficients for all 168 pairwise combinations of class and test--effort metrics.
For this research question, we are not interested in the direction of the correlation (positive or negative), and thus we use absolute values $|\rho|$. The colors range from blue (maximum correlation, $|\rho| = 1.0$) to red (minimum correlation, $|\rho| = 0.0$). We interpret $|\rho|$ as \emph{weak} ($\leq 0.3$), \emph{moderate} ($0.3$--$0.5$), or \emph{strong} ($\geq 0.5$), following the widely accepted classification of Cohen~\cite{cohen2013statistical}.

\begin{figure}[h]
  \centering
  \includegraphics[width=\linewidth]{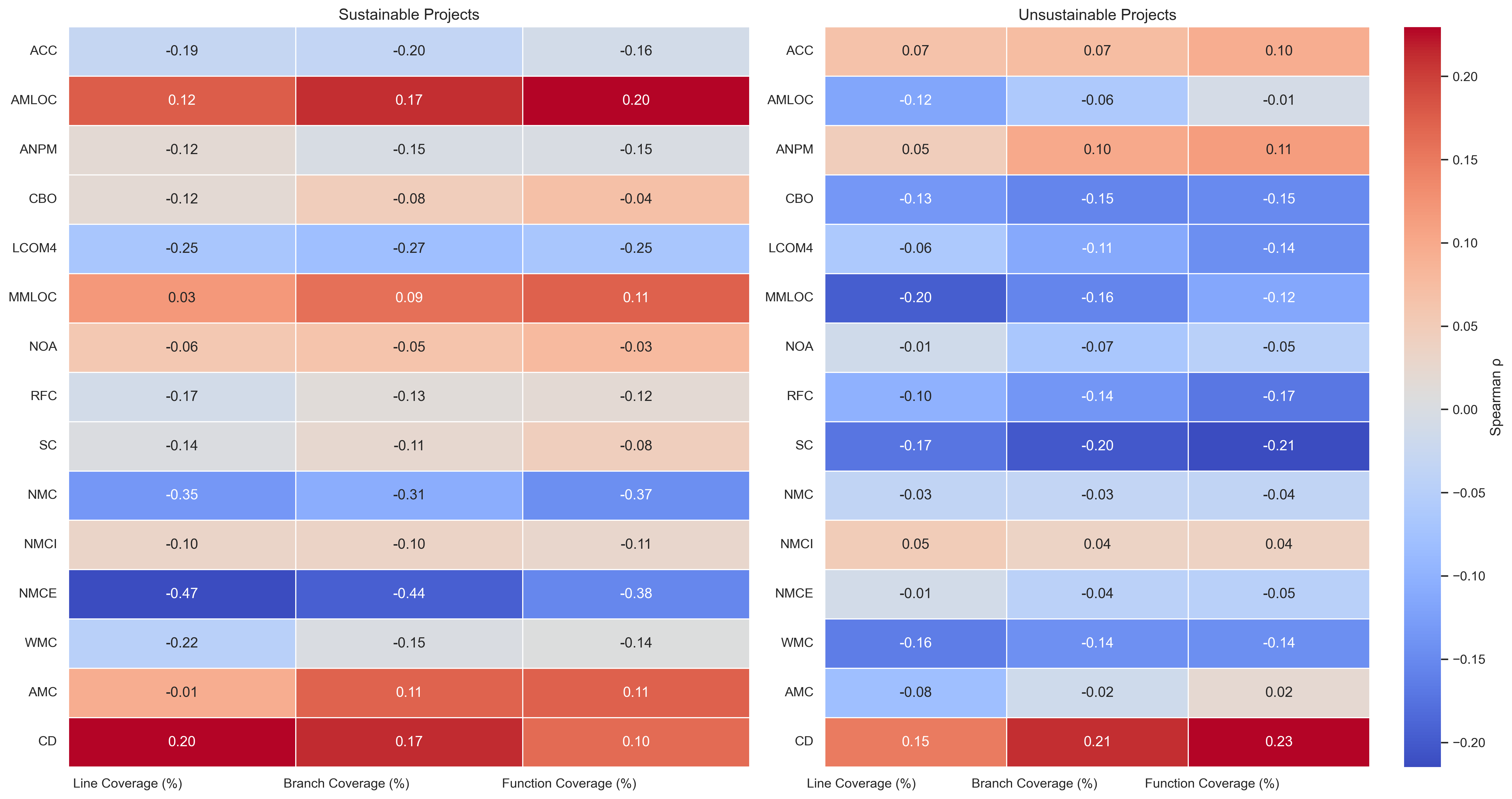}
     \caption{Comparison of heatmaps for sustainable and unsustainable projects using $t$-year sustainability classification.}
   \label{fig:heatmap_comparison_tyear}
\end{figure}

The heatmap in Figure~\ref{fig:heatmap_comparison_tyear}  illustrates the Spearman correlation ($\rho$) between static code metrics and test coverage metrics (line, branch, and function) across sustainable and unsustainable software projects. In sustainable projects, stronger and more consistent correlations are observed. Notably, NMCE and NMC show moderate to strong negative correlations with all types of coverage, while LCOM4 and WMC also correlate negatively, suggesting that high complexity and low cohesion are associated with reduced test coverage. In contrast, CD and AMLOC show weak to moderate positive correlations, implying better documentation and modularity may or may not improve testability. On the other hand, unsustainable projects show generally weaker and more erratic correlations, with most values near zero. Although CD retains a positive trend, the absence of strong relationships suggests a lack of consistent structural-testability patterns in unsustainable projects' codebase.

\begin{figure}[h]
  \centering
  \includegraphics[width=\linewidth]{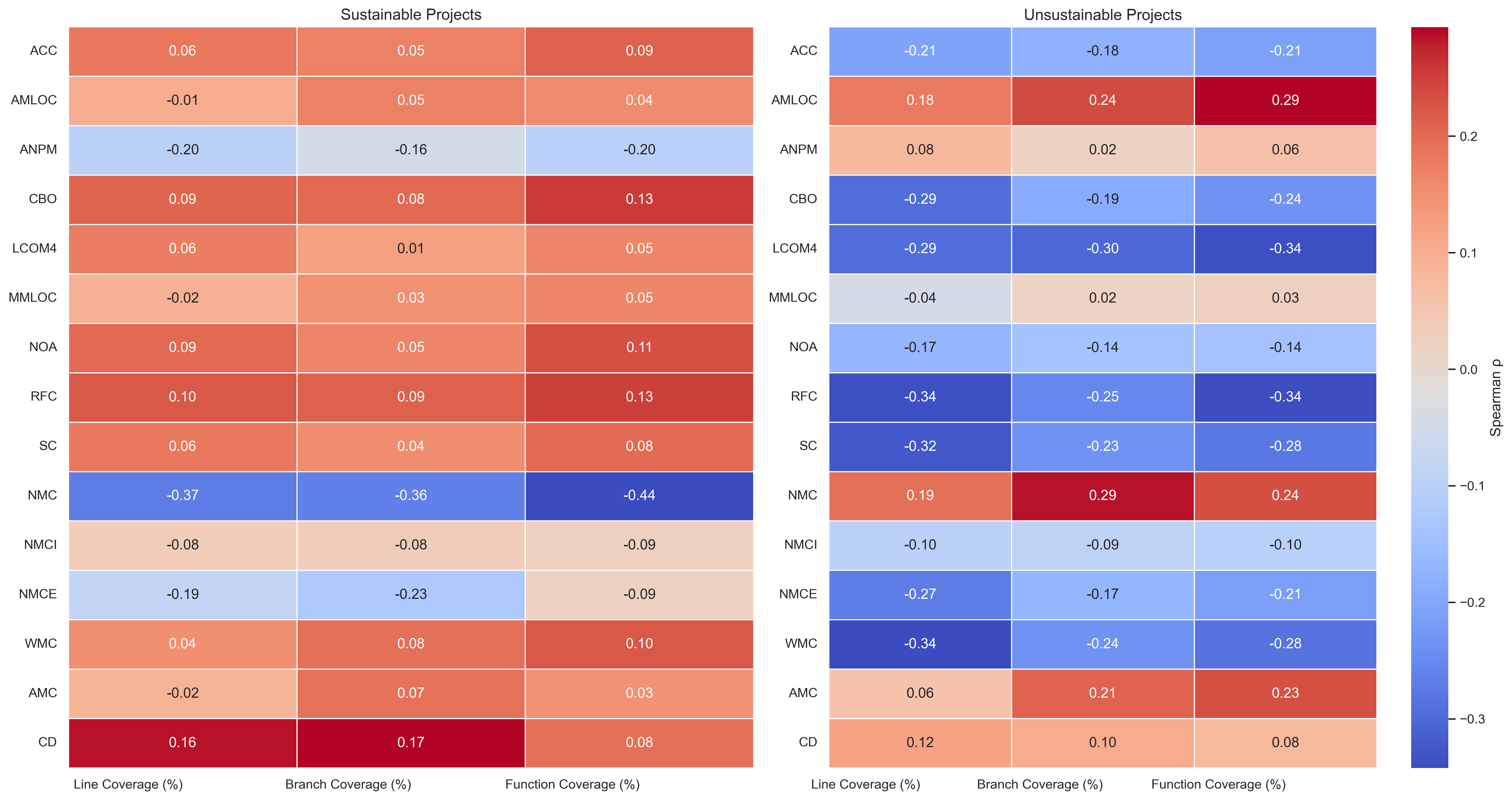}
     \caption{Comparison of heatmaps for sustainable and unsustainable projects using TF classification.}
   \label{fig:heatmap_comparison_tf}
\end{figure}

Figure~\ref{fig:heatmap_comparison_tf} presents the same comparison, but for classification based on TF. 
% The side-by-side heatmap reveals notable differences in how code metrics correlate with test coverage in sustainable versus unsustainable projects. 
In sustainable projects, metrics like NMC and ANPM show strong negative correlations with coverage, suggesting that increased structural complexity is associated with poorer testability. Conversely, CD consistently exhibits a positive correlation, indicating that better-documented code tends to have higher coverage. In contrast, unsustainable projects display weaker or even reversed correlations---for instance, NMC and AMC (Average Method Complexity) correlate positively with coverage, which may reflect inconsistent or inflated metric distributions in projects lacking long-term maintenance. The metric RFC, positively correlated with coverage in sustainable projects, turns negative in unsustainable ones, underscoring a possible breakdown in cohesion between code structure and testability. Overall, sustainable projects exhibit more coherent and interpretable correlations, while unsustainable ones demonstrate erratic or contradictory patterns, reinforcing the idea that well-maintained codebases tend to follow better engineering and testing practices.

\begin{boxC}
    In both classification schemes, sustainable projects show stronger and more consistent correlations between code structure and test coverage, particularly with complexity and documentation metrics. Unsustainable projects, by contrast, exhibit weaker or inconsistent relationships, suggesting less disciplined testing practices and structural design.
\end{boxC}

We further investigate whether different metrics can help distinguish between sustainable and unsustainable projects, aiming to understand if such metrics offer insights into project longevity. To assess these differences, we perform two-sided Mann–Whitney U tests and visually examine the distributions. To account for the multiple comparisons across various metrics, we apply the Benjamini-Hochberg procedure~\cite{benjamini1995controlling}, which offers greater statistical power compared to other correction methods. Additionally, we compute effect sizes using Cliff’s delta~\cite{grissom2005effect}, interpreted according to the thresholds proposed by Romano et al.~\cite{romano2006appropriate}.
% Figure 12 shows violin plots with the distributions of the number of developers, commits and files, and project age measured in days of the surviving and non-surviving projects. All values refer to the date of the studied TFDDs.
\begin{figure}[h]
  \centering
  \includegraphics[width=.8\linewidth]{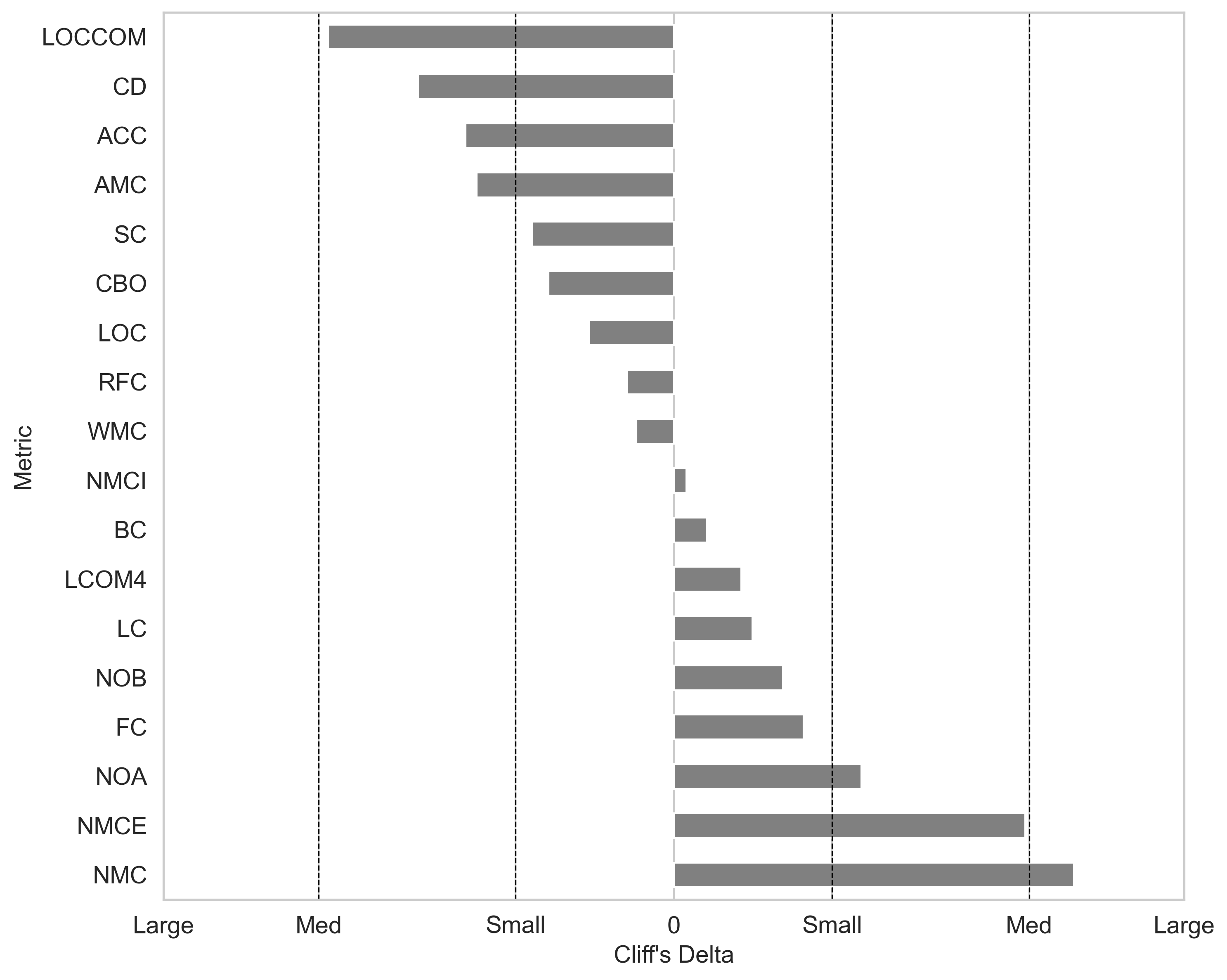}
     \caption{Cliff's Delta on $t$-year sustainability classification.}
   \label{fig:tyear_cd}
\end{figure}

The Cliff’s Delta plot comparing software metrics between sustainable and unsustainable projects (based on \emph{t}-year sustained activity) in Figure~\ref{fig:tyear_cd} shows that sustainable projects tend to have higher modularity and test coverage (NMC, NMCE, NOA, FC) and slightly more function and branch coverage (LC, BC), as indicated by positive effect sizes. In contrast, unsustainable projects exhibit significantly higher values in documentation-related metrics (LOCCOM, CD), coupling and complexity measures (ACC, AMC, CBO, SC), and overall size (LOC, RFC, WMC). Metrics like LCOM4, NOB, and NMCI show negligible differences, suggesting they are less predictive of sustainability.

\begin{figure}[h]
  \centering
  \includegraphics[width=.8\linewidth]{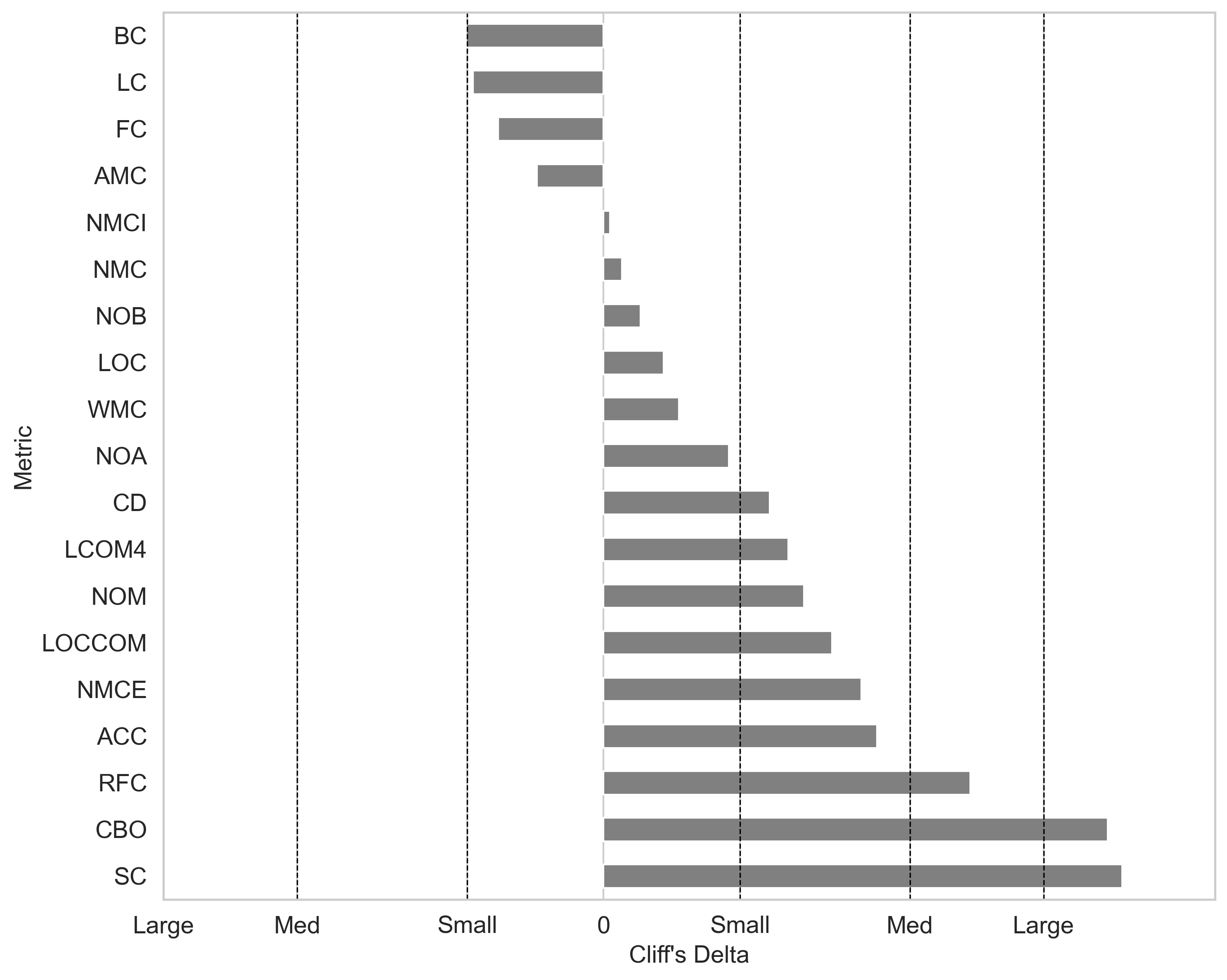}
     \caption{Cliff's Delta on TF sustainability classification.}
   \label{fig:tf_cd}
\end{figure}

Figure~\ref{fig:tf_cd} presents the Cliff's Delta effect sizes for various code and testability metrics when comparing sustainable and unsustainable projects. Positive values indicate higher metric values in sustainable projects, while negative values indicate the opposite. The results show that sustainable projects exhibit substantially higher structural complexity (SC), coupling (CBO, ACC), class-level response (RFC), external function interaction (NMCE), and comment-related metrics (LOCCOM, CD), with medium to large effect sizes. This suggests that sustainable projects tend to be more complex, interconnected, and better documented. In contrast, test coverage metrics, such as function (FC), line (LC), and branch coverage (BC), are slightly higher in unsustainable projects, though the effect sizes are small. This may reflect simpler codebases or limited feature evolution in unsustainable projects. Overall, sustainable projects demonstrate more mature and intricate code structures, while unsustainable ones show marginally better test coverage.
\begin{boxF}{Result}
Sustainable projects tend to exhibit higher structural complexity, better documentation, greater modularity, and more consistent correlations between code metrics and test coverage. While projects classified as sustainable under the \emph{t}-year criterion show higher test coverage, those deemed unsustainable by the TF-based method display slightly better coverage, likely due to concentrated early development. Overall, mature architecture, modular design, and strong documentation, not coverage alone, serve as more reliable indicators of long-term sustainability.
\end{boxF}
\subsection{RQ2: Does using a stricter threshold for sustained activity improve the reliability of sustainability classifications based on test quality metrics?}

Despite observing meaningful correlations between structural metrics and test coverage in sustainable projects (as defined in RQ1), the classification method significantly influenced the interpretation of test coverage trends. According to the t-year criterion, which requires long-term development activity, sustainable projects consistently exhibited higher test coverage across all metrics (line, function, and branch), suggesting that enduring maintenance often coincides with better testing practices. In contrast, the TF-based classification, which focuses on developer continuity, yielded the opposite trend: unsustainable projects showed higher test coverage. This apparent contradiction implies that test coverage alone may not be a definitive signal of sustainability.

To further investigate this assumption, we apply a stricter threshold for sustainability classification, following the methodology proposed by Xiao et al.\cite{xiao2023early}. In their study, the authors develop a machine learning model to predict the long-term sustainability of open-source GitHub projects based on early behavioral signals. They define a project as sustainable if it remains active for at least two years ($t = 2$) and maintains a median number of commits per month greater than or equal to a threshold $k$, where $k$ controls the strictness of the sustainability label. In Section 4.3 (Sensitivity Analysis), they examine the impact of increasing $k$ (e.g., from 1 to 2 to 6) on classification performance and find that stricter thresholds improve model accuracy, as reflected in higher AUC and F1 scores. These findings support the idea that a more rigorous sustainability definition reduces label noise and leads to more reliable distinctions between sustainable and unsustainable projects. Motivated by this, we conducted additional experiments using increasingly strict $k$-values to refine our sustainability labels and test whether the observed inconsistencies in coverage trends persist under more robust classification criteria.

\begin{figure}[htbp]
    \centering

    \begin{subfigure}[t]{\linewidth}
         \includegraphics[width=\linewidth]{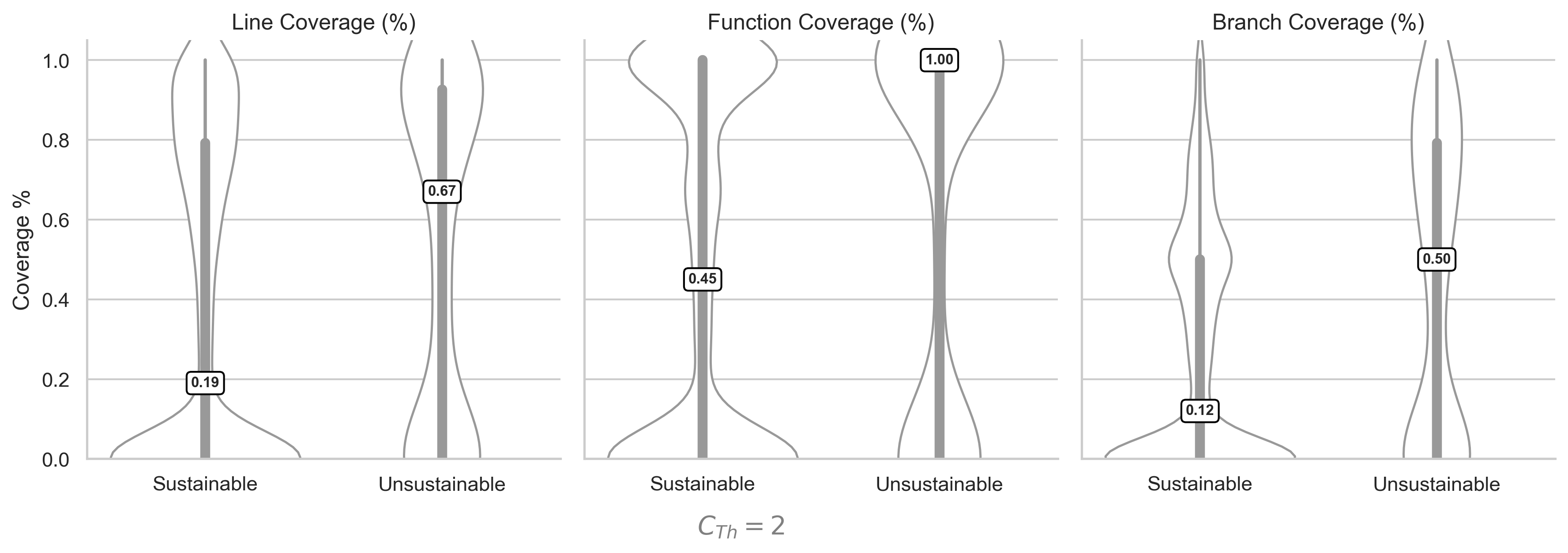}
       
        \label{fig:freq2}
    \end{subfigure}

    \begin{subfigure}[t]{\linewidth}
        \includegraphics[width=\linewidth]{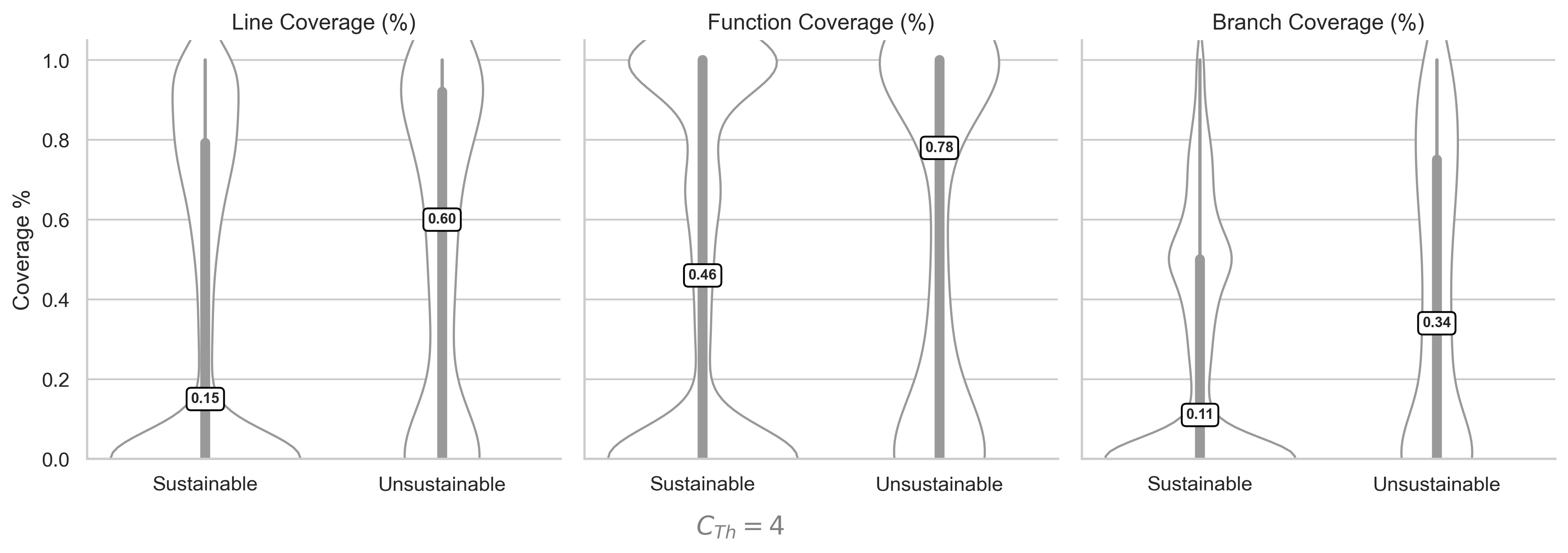}
       
        \label{fig:freq4}
    \end{subfigure}

    \begin{subfigure}[t]{\linewidth}
        \includegraphics[width=\linewidth]{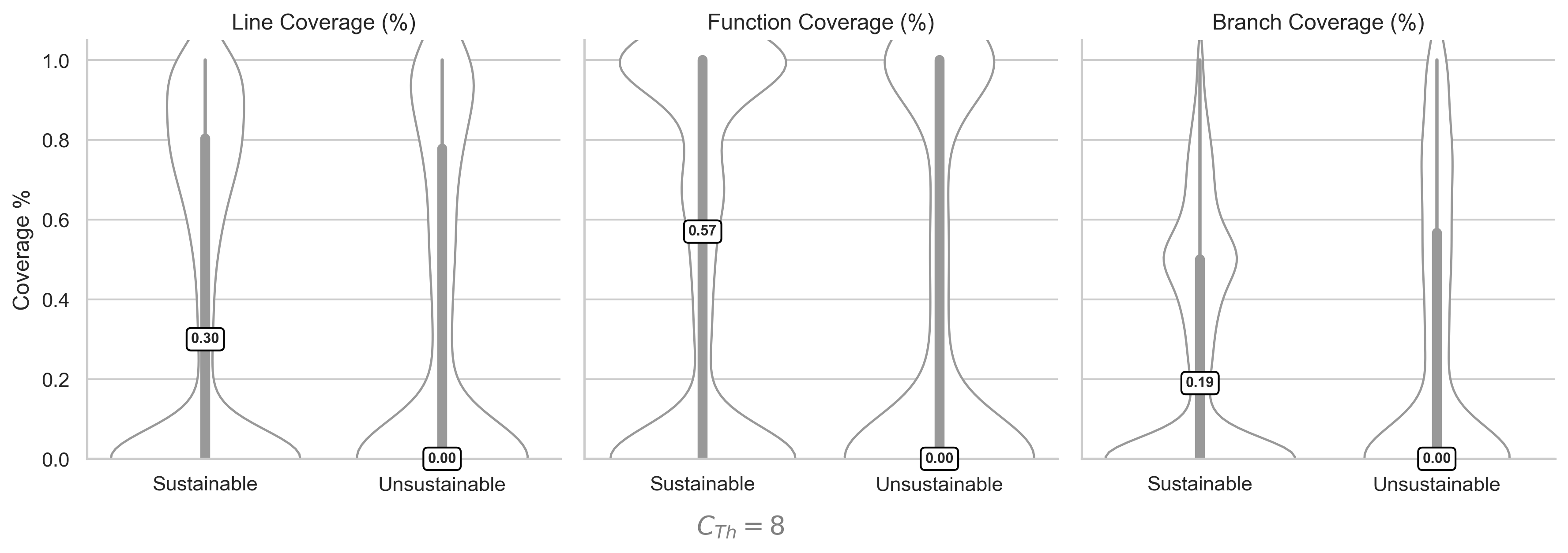}
       
        \label{fig:freq8}
    \end{subfigure}

    \begin{subfigure}[t]{\linewidth}
        \includegraphics[width=\linewidth]{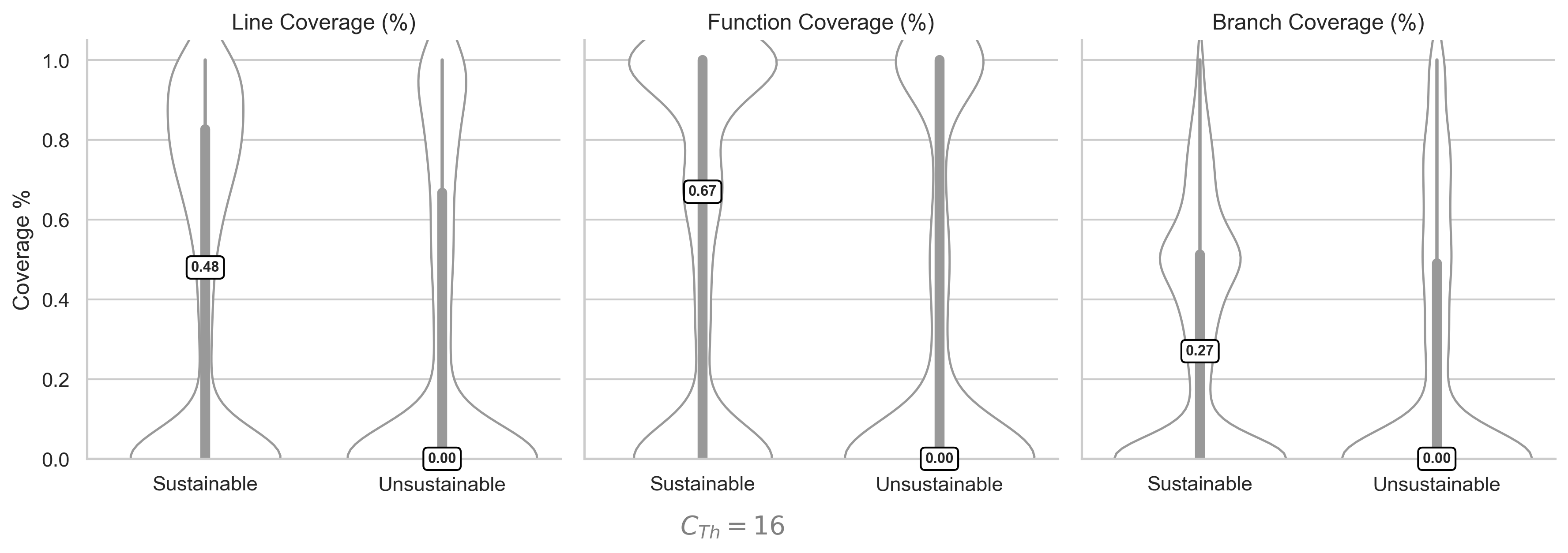}
       
        \label{fig:freq16}
    \end{subfigure}

        \begin{subfigure}[t]{\linewidth}
        \includegraphics[width=\linewidth]{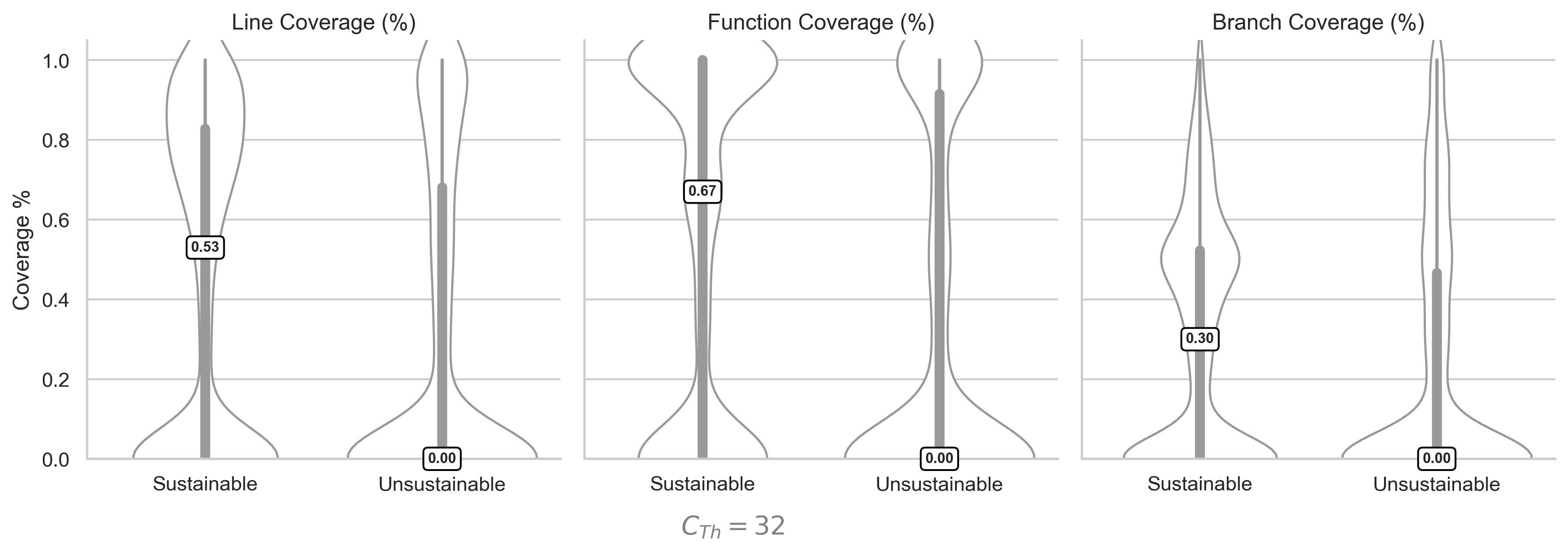}
       
        \label{fig:freq32}
    \end{subfigure}

    \caption{Coverage plots across increasing commit frequency thresholds.}
    \label{fig:frequency_coverage_all}
\end{figure}

The series of violin plots from Figure~\ref{fig:frequency_coverage_all} illustrates how test coverage varies between sustainable and unsustainable projects under increasing thresholds of commit frequency. At lower thresholds (e.g., 2 and 4 commits per month), unsustainable projects surprisingly exhibit higher median coverage values, suggesting that some dormant or less active projects were once well-tested. However, as the threshold becomes more stringent (e.g., 8, 16, and 32), this pattern reverses: sustainable projects show consistently higher median coverage across all three dimensions—line, function, and branch—while unsustainable projects drop to near-zero coverage. This shift indicates that truly active and frequently maintained projects are more likely to invest in and maintain systematic testing infrastructure. Thus, under stricter definitions of sustained activity as a proxy of sustainability, test coverage becomes a more reliable indicator of sustained project health and engineering discipline.

\begin{figure}[h]
  \centering
  \includegraphics[width=\linewidth]{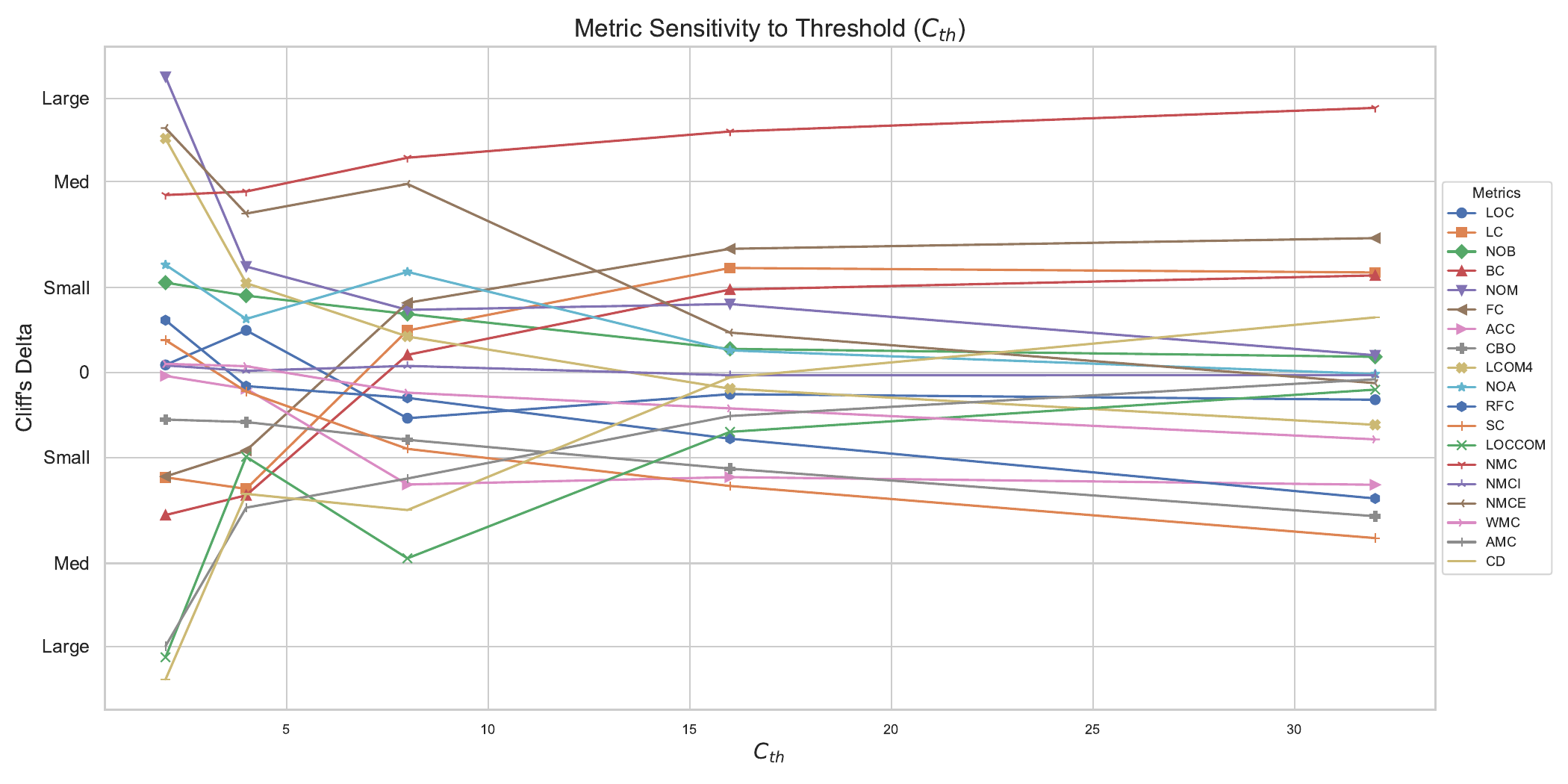}
     \caption{Comparing Cliff's delta by changing threshold.}
   \label{fig:thershold_metrics}
\end{figure}

Figure~\ref{fig:thershold_metrics} presents Cliff's Delta values across different sustainability thresholds ($C_{\text{Th}}$), comparing sustainable and unsustainable projects. Across all thresholds, sustainable projects consistently show higher values in metrics related to testing and documentation, such as FC, LC, BC, CD, and LOCCOM, indicating a stronger emphasis on code quality and maintainability. In contrast, structural and complexity-related metrics like SC, WMC, CBO, and RFC tend to be higher in unsustainable projects, suggesting greater code complexity and interdependence. Notably, the NMC is substantially higher in sustainable projects, especially at higher thresholds, reflecting their modular and reusable design practices. As the threshold decreases, the effect sizes of structural metrics diminish, while the positive association between coverage, documentation, and sustainability remains robust.

\begin{boxF}{Result}
% Initial inconsistencies in test coverage trends—where TF-based unsustainable projects showed higher coverage—prompted a deeper analysis using stricter sustainability thresholds based on commit frequency. As the threshold ($k$) increased, sustainable projects consistently exhibited higher median coverage across all metrics, while unsustainable projects dropped to near-zero coverage. This observation confirms that stricter activity-based definitions more effectively capture sustainable behavior. Additionally, an effect-size analysis shows that sustainable projects emphasize modularity, test coverage, and documentation, while unsustainable ones exhibit higher structural complexity and coupling. These findings reinforce that under refined criteria, coverage and quality-focused metrics serve as stronger indicators of sustainability.
Initial inconsistencies in TF-based results, where some unsustainable projects showed higher coverage, motivated a re-evaluation using stricter thresholds. As the commit-frequency threshold ($k$) increased, sustainable projects showed higher median coverage, while unsustainable projects dropped toward zero. This confirms that stricter activity-based definitions better capture sustainable behavior. Effect-size analyses further show that sustainable projects favor modularity, documentation, and higher coverage, whereas unsustainable ones exhibit greater complexity and coupling. Overall, refined criteria strengthen the link between code quality and sustainability.
\end{boxF}

\subsection{RQ3: Is there any strong relationship between code metrics and test coverage in scientific software overall?}
\begin{figure}[h]
  \centering
  \includegraphics[width=\linewidth]{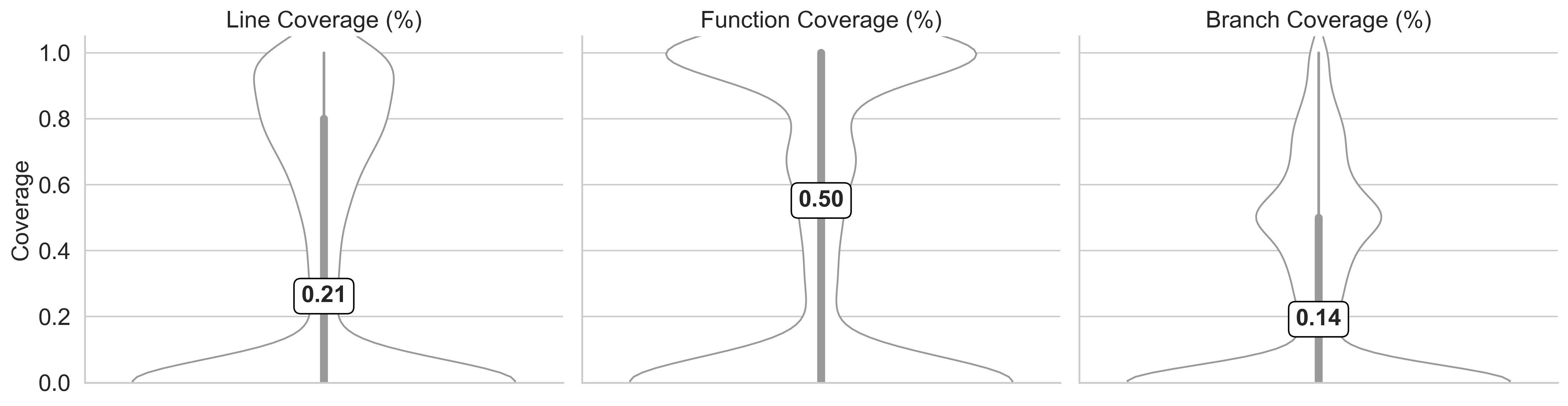}
     \caption{Test Coverage of the projects.}
   \label{fig:overall_coverage_boxchart}
\end{figure}
The violin plots in Figure~\ref{fig:overall_coverage_boxchart} illustrate the distribution of test coverage metrics—line, function, and branch coverage—across all projects. The median values reveal generally low coverage, with line coverage at 21\%, function coverage at 50\%, and branch coverage at just 14\%. While function coverage appears relatively higher, the distributions for both line and branch coverage are heavily skewed toward zero, indicating that many projects have minimal or no test coverage. The function coverage distribution is notably bimodal, suggesting a divide between well-tested and poorly-tested projects. Overall, these results highlight inconsistent testing practices and suggest that significant improvements are needed, particularly in achieving adequate line and branch coverage.

\begin{figure}[h]
  \centering
  \includegraphics[width=.8\linewidth]{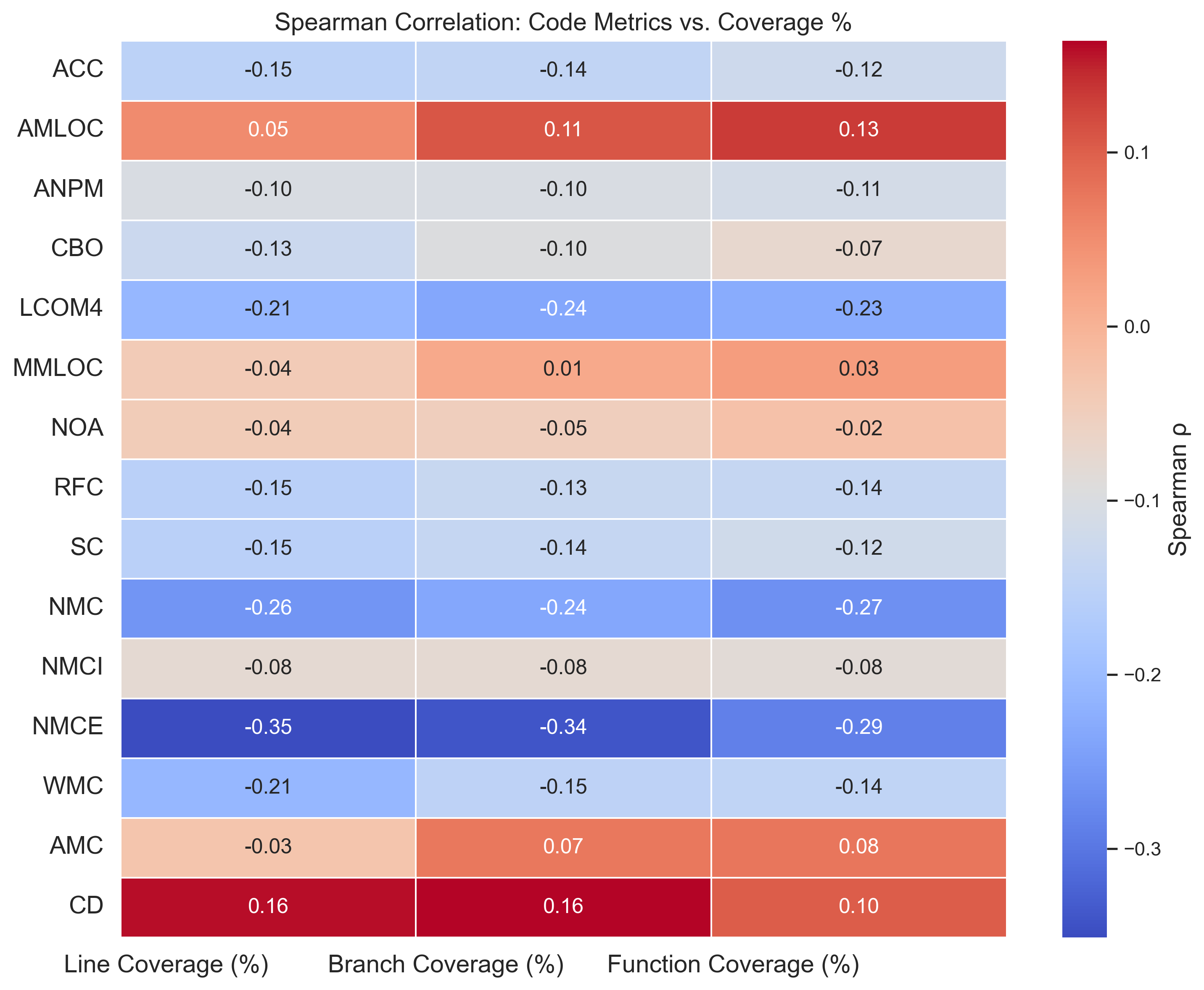}
     \caption{Heatmap for the full dataset.}
   \label{fig:overall_heatmal}
\end{figure}

The heatmap in Figure~\ref{fig:overall_heatmal} presents the Spearman correlation coefficients between various code metrics and three coverage metrics: line, branch, and function coverage percentages. Overall, most code metrics exhibit weak negative correlations with test coverage, suggesting that as certain structural or complexity-related attributes of the code increase, coverage tends to decrease. Notably, the strongest negative correlations are observed with NMCE, which shows moderate negative associations with all three coverage types (e.g., $\rho$ = -0.35 with line coverage), indicating that code with more external dependencies may be harder to test thoroughly. Similarly, NMC also shows moderately negative correlations across the board. Metrics such as LCOM4, WMC, and RFC, which reflect cohesion and complexity, also have modest negative correlations. In contrast, metrics like AMLOC and CD show slight positive correlations, possibly indicating that more verbose and well-documented code may facilitate better testability. However, these correlations are weak and should be interpreted cautiously. The overall trend supports the notion that higher code complexity and coupling are generally associated with lower test coverage across scientific software projects.
\begin{boxF}{Result}
    Overall, test coverage in scientific software is low and uneven, with many projects lacking adequate line and branch coverage. Correlation analysis shows that higher code complexity and coupling (e.g., NMCE, NMC) are moderately associated with lower test coverage, while documentation and verbosity show slight positive links. These findings suggest that structural complexity hampers testability across the board.
\end{boxF}

\section{Discussion}
Scientific software sometimes exhibits intricate, non-deterministic behavior with numerous execution paths and extensive input demands, making it challenging to manually define key input boundaries or effectively partition the input space for testing purposes~\cite{eisty2020testing}. Developers of such software often have a limited understanding of standard software engineering testing practices and may prioritize producing results over ensuring software quality, driven by budget constraints or the exploratory nature of research~\cite{kanewala2014testing}. Moreover, many lack formal training in software engineering, treating testing as a secondary activity~\cite{segal2007some}. Our experimental observations are consistent with these findings: the scientific software under study demonstrates poor test coverage in its existing test suite, underscoring the persistent difficulties in achieving robust testing for research-oriented codebases.

Beyond testing challenges, structural and cultural issues further hinder sustainability. Software citation is not yet a standard practice in research workflows; software is often not published in a citable form, and metadata quality and persistence remain unreliable. Support from publishers and funders is limited, software is not consistently recognized as a valued research output, credit attribution is ambiguous, and dependencies are frequently overlooked. These technical and social challenges reinforce one another, impeding proper maintenance and testability~\cite{druskat2021research}.

Overall, our analysis reveals that while sustainable projects tend to achieve higher test coverage under time-based sustainability measures (t-year criterion), this trend does not hold under the TF-based classification, indicating that high coverage alone is not a sufficient marker of long-term project health. Across both classification methods, sustainable projects consistently demonstrate stronger correlations between structural quality metrics, such as complexity, modularity, and documentation, and test coverage. This observation suggests that these projects adopt more disciplined engineering and testing practices. Conversely, unsustainable projects show weaker or inconsistent relationships, implying less structured development and testing efforts. When stricter activity-based sustainability thresholds are applied, sustainable projects maintain higher coverage and structural integrity, while unsustainable ones show a significant decline, confirming that ongoing development activity better reflects sustainability. Overall, the study emphasizes that true software sustainability is rooted not merely in test coverage but in robust architecture, strong documentation, and consistent maintenance practices that promote long-term viability and adaptability.

In addition, our investigation of projects from the CASS Portfolio revealed that a significant number lacked proper containerization and updated build documentation. This finding corroborates prior research showing that research software often suffers from poor reproducibility, portability, and deployment support, which are factors that critically undermine long-term sustainability~\cite{nyenah2024software}.

\subsection{RQ1: Practical Implications}

The findings from RQ1 suggest that sustainable projects exhibit stronger relationships between code structure and testing practices, along with more consistent quality patterns. Practically, this implies that developers should not treat testing and code design as separate concerns. Instead, they should adopt integrated engineering practices where modularity, low coupling, and clear structure are prioritized alongside test development. For project maintainers and funding agencies, these results highlight the value of monitoring internal code metrics (e.g., complexity, cohesion, coverage trends) as early warning signals of declining sustainability. Tooling support can also be developed to automatically flag projects with poor structural-test alignment, enabling proactive intervention before sustainability issues become critical.

\subsection{RQ2: Practical Implications}

The results of RQ2 demonstrate that stricter activity-based definitions of sustainability lead to more consistent and reliable interpretations of quality metrics. This has important implications for both researchers and practitioners. For researchers, it emphasizes the need to carefully select and justify sustainability definitions to avoid misleading conclusions. For practitioners and policymakers (e.g., funding bodies, open-source governance organizations), it suggests that simple activity thresholds (such as commit frequency) can serve as practical proxies for identifying genuinely sustainable projects. Additionally, project evaluation frameworks should incorporate stricter criteria to better distinguish between actively maintained and superficially active or legacy systems. This can improve decision-making in resource allocation, project adoption, and long-term support strategies.

\subsection{RQ3: Practical Implications}

The findings from RQ3 indicate that higher code complexity and coupling are generally associated with lower test coverage, while better documentation and modularity may support improved testability. This has direct implications for software engineering practices in scientific domains. Developers should actively manage complexity and reduce interdependencies to make their code more testable. Practices such as modular design, refactoring, and improved documentation can lower the barriers to writing effective tests. For tool builders, these insights open opportunities to create automated tools that predict testability issues based on static code metrics, enabling early intervention. For educators and research institutions, the results reinforce the importance of training scientists in software engineering best practices, particularly in designing testable and maintainable code.

% Sustainable proejcts has more function coverage, BC, LC, NMC (always medium high in Sustainable projects). SC lowers.~\ref{fig:thershold_metrics}

% Unlike previous research to find correlationship among metrics from OSS projects~\cite{terragni2020measuring}, Most of the SOSS code metrics shows no correlation with test efforts as is shown by Figure~\ref{fig:overall_heatmal}. Even, Figure~\ref{fig:heatmap_comparison_tf} and ~\ref{fig:heatmap_comparison_tyear} shows majority of the metrics has no correlation with test effort when the dataset is divided into two groups. Only, NMC, NMCE shows some negative correlation with test efforts in sustainable projects (in overall dataset as well). 
% Which means that Sustainable projects are more modular than unsustainable SOSS projects. 

%Mention about legacy code and lack of understanding.
%Difficult to test (because of test oracle problem).
%

\section{Threats to Validity}
\textbf{\textit{Construct Threats.}}
The construct threat pertains to how sustainability is defined for the SciOSS. In this study, we derived the sustainability labels from prior research and validated them against the literature~\cite {valiev2018ecosystem,xiao2023early,avelino2019abandonment,han2024sustainability}. Sustained development activity and developer continuity are used as operational proxies for sustainability. Both are supported by prior work as reliable indicators of long-term maintenance and engagement. By employing two complementary classification methods validated in prior studies, we ensure a multidimensional perspective and reduce construct bias, thereby enhancing the validity of the sustainability construct.

\textbf{\textit{External Threats.}}
External threats concern the generalizability of the findings. The dataset is drawn from the CORSA catalog, which focuses on high-quality scientific software projects written in C/C++. While this ensures strong relevance to scientific computing, it limits applicability to other ecosystems, such as Python-based data analysis or Java enterprise software. Moreover, our dataset includes only 14 projects—the subset that could be successfully built, instrumented for coverage, and analyzed with a uniform metric-extraction pipeline. This necessarily excludes projects with complex MPI/GPU configurations, specialized hardware requirements, missing or non-runnable tests, or build systems that resist reproducible compilation. As a result, the dataset overrepresents well-engineered, actively maintained projects and underrepresents highly experimental, abandoned, or hard-to-build codebases, introducing survivorship and instrumentation bias common in empirical studies of scientific software. While these constraints limit broad generalizability, they ensure consistent, reproducible metric collection, which is essential for structural and test-quality analysis. Future work will expand CORSA and explore techniques for instrumenting more complex or hardware-dependent projects.

% Additionally, only projects that could be successfully built and instrumented for coverage were included, which may exclude highly experimental, abandoned, or legacy codebases. As a result, the conclusions may not fully extend to the broader software landscape.

\textbf{\textit{Internal Threats.}}
Internal threats relate to the robustness of data collection and statistical inference. We used non-parametric statistical tests (Mann–Whitney U, Spearman’s rank, and Cliff’s delta) due to the non-normality of metric distributions, verified with the D’Agostino $k^2$ test. The Benjamini–Hochberg correction controls for false discovery in multiple comparisons. However, the small sample size (14 projects) limits statistical power, making it harder to detect small effects and increasing uncertainty in marginal results. Despite rigorous testing, the limited dataset constrains internal reliability.

\textbf{\textit{Conclusion Threats.}}
To mitigate measurement and instrumentation bias, the study uses a standardized, tool-agnostic metric extraction pipeline centered on Analizo~\cite{terceiro2010analizo}. This framework normalizes both procedural and object-oriented C/C++ projects by modeling each translation unit as a class-like module, ensuring consistent computation of structural metrics (e.g., CBO, LCOM4, WMC, NOM). The use of complementary tools, such as Lizard, Cloc, and custom Clang AST analyzers, further strengthens metric accuracy and coverage across complexity, documentation, and function calls. These design choices enhance measurement consistency and, consequently, the reliability of the study’s conclusions.

We also acknowledge limitations in the structural metrics derived for this study. OO-based measures such as LCOM4, CBO, and RFC were developed for class-oriented systems and only approximate cohesion and coupling in procedural C, where they largely capture patterns of global-state sharing or inter-function interaction rather than classical OO semantics. Likewise, our call metrics (NMC, NMCI, NMCE) rely on per–translation unit Clang AST analysis, which under-approximates call relationships involving function pointers, macro expansions, templated or inlined code, and cross-unit interactions—well-known challenges in static call-graph construction for C/C++. Accordingly, we interpret both cohesion/coupling and call metrics as conservative, relative indicators of structural organization rather than complete or absolute measures of modular quality.

\section{Conclusion}
Sustaining SciOSS is critical for ensuring the reproducibility, reliability, and long-term impact of computational research. This study investigated the relationship between internal software quality and the sustainability of SOSS projects. By analyzing C/C++ projects from the CORSA Catalog, we classified projects using two validated methods and compared their structural and test quality characteristics. Our results show that sustainable projects tend to be better tested, more modular, and less complex than unsustainable ones. Additionally, stricter sustainability thresholds yield more consistent trends in quality metrics, suggesting that refined definitions can enhance classification reliability. We also find that several code metrics are moderately correlated with test coverage, indicating that structural properties may serve as proxies for testing effort and software maintainability. These findings highlight the importance of incorporating internal quality assessments into sustainability evaluations and can inform future efforts to support the development of robust, maintainable, and enduring scientific software.

\section{Acknowledgment \& Data}
We acknowledge the support of the Consortium for Open-Source Research Software Advancement (CORSA), a project supported by the U.S. Department of Energy, Office of Science, Office of Advanced Scientific Computing Research, Next-Generation Scientific Software Technologies program, under contract number DE-AC05-00OR22725. 
\label{data_avail}
We share all code and data used in this study at \textcolor{blue}{\url{https://github.com/Mushfiq344-utk/SciOSS_Sustainability}}.

\bibliographystyle{ACM-Reference-Format}
\bibliography{sample-base}

\appendix

\end{document}